\documentclass[twocolumn,showpacs,preprintnumbers,amsmath,amssymb]{revtex4}
\usepackage{graphicx}
\usepackage{dcolumn}
\usepackage{bm}
\newcommand{\beq}{\begin{equation}}
\newcommand{\enq}{\end{equation}}

\begin{document}
\title {Nonexistence of "Spin Transverse Force" for a Relativistic
Electron}
\author{ Wlodek Zawadzki}
\email{zawad@ifpan.edu.pl}
 \affiliation{ Institute of Physics, Polish Academy of
Sciences, 02668 Warsaw, Poland}
 \begin{abstract}
  Using the complete Dirac Hamiltonian for a relativistic electron it is shown that
 the spin transverse force derived by S.Q. Shen, Phys. Rev. Lett.
 {\bf 95}, 187203 (2005) does not exist. This force is an artefact resulting
 from an approximate form of the employed Hamiltonian.
\end{abstract}
\maketitle
 \indent In a recent paper Shen [1] considered a force acting on a
 relativistic electron in the presence of an external electric
 potential and a magnetic field. Using an approximate Hamiltonian,
 which contained the spin-orbit interaction and the Darwin term,
 it was shown that the electric field exerts a transverse force on
 an electron spin when the electron is moving. The effects
 resulting from this force in semiconductors were also considered.

 In the following we critically reexamine this problem and show
 that the spin transverse force does not exist. To begin with, the
 Hamiltonian used in Ref. 1 is incomplete. As is well known, in the
 $(v/c)^2$
 approximation to the Dirac equation there appear three terms in addition to the
 Pauli equation: the spin-orbit term, the Darwin term and
 the $(\textbf{p}-e\textbf{A})^4$ term, in which $\textbf{p}$ is the momentum and $\textbf{A}$ is the vector
 potential of a magnetic field. It was recently shown that the last
 contribution should be completed by a spin term, see Ref. 2. The
 last two terms, which are also of the $(v/c)^2$ order, were omitted in
 [1]. Second, the author used for the force the formula $\textbf{F}=m_\text{0}d\textbf{v}/dt$   ,
 where $m_\text{0}$ is the rest electron mass. This is incorrect since in
 the relativistic range the correct formula for the force is $\textbf{F}=dm(v)\textbf{v}/dt$  ,
 where $m(v)$ is the velocity-dependent Lorentz mass. However, the essential point is that, in order to calculate the
 force, the author used an approximation to the Dirac Hamiltonian,
 while an exact solution to this problem exists and gives a
 different result, see Strange [3]. Below we reproduce this
 derivation with some comments.

  The complete Dirac Hamiltonian with the electric and magnetic
 potentials is, in the standard notation,
\begin{equation}
H=c\bm{\alpha}\cdot (\textbf{p}-e\textbf{A})+\beta m_0 c^2+V \;\;,
\end{equation}
where the potential energy is $V=e\phi(\textbf{r})$, the magnetic
field is
 $\textbf{B}=\bm{\nabla} \times \textbf{A}(\textbf{r})$,
 and $\bm{\alpha}_i$
 are the Dirac matrices. The velocity is
\begin{equation}
\textbf{v}=\frac{d\textbf{r}}{dt}=\frac{i}{\hbar}[H,\textbf{r}]=c\bm{\alpha}\;\;.
\end{equation}
The force is
\begin{equation}
\textbf{F}=\frac{d}{dt}(\textbf{p}-e\textbf{A})\;\;.
\end{equation}
 The force is rarely written this way, but the above
 formula is correct since $\textbf{p}-e\textbf{A} = m(v)\textbf{v}$, see for example Refs. 4,5.
 We have
\begin{equation}
\frac{d\textbf{p}}{dt}=\frac{i}{\hbar}[H,\textbf{p}]=-\bm{\nabla}V+
ec\bm{\nabla}(\bm{\alpha} \cdot \textbf{A})\;\;,
\end{equation}
\begin{equation}
\frac{d\textbf{A}}{dt}=\frac{\delta\textbf{A}}{\delta
t}+\frac{i}{\hbar}[H,\textbf{A}]= \frac{\delta\textbf{A}}{\delta
t}+c(\bm{\alpha} \cdot \bm{\nabla})\textbf{A}\;\;,
\end{equation}
where $\nabla$ is the differentiation vector acting only on $V$ or
$\textbf{A}$, not on the wave function. Hence
\begin{equation}
\textbf{F}=e(-\bm{\nabla}\phi-\frac{d\textbf{A}}{dt})+ec[\bm{\nabla}(\bm{\alpha}\cdot
\textbf{A})-(\bm{\alpha}\cdot\bm{\nabla})\textbf{A}]\;\;.
\end{equation}
Since $\bm{\nabla}(\bm{\alpha}\cdot\textbf{A})-(\bm{\nabla} \cdot
\alpha)\textbf{A})=
\bm{\alpha}\times(\bm{\nabla}\times\textbf{A})$, we finally obtain
\begin{equation}
\textbf{F}=e\textbf{E}+e\textbf{v}\times\textbf{B}\;\;,
\end{equation}
 where we have used the above formulas for $\textbf{v}$ and $\textbf{B}$ and the standard
 relation for an electric field $\textbf{E}=-\bm{\nabla}\phi-\delta \textbf{A}/\delta t$.   .

 Thus the classical result for the Lorentz force is obtained,
 apart from the fact that $\textbf{v}$ is an operator. In particular, once the
 force is expressed by the velocity (as it is in Ref. 1), the effects
 of the spin-orbit interaction do not appear. It should be
 mentioned that in the non-relativistic quantum mechanics one
 should use the symmetrized product $(\textbf{v}\times\textbf{B}-\textbf{B}
 \times\textbf{v})/2$
  since the velocity $\textbf{v}=\textbf{p}/m_0$
 does not commute with $\textbf{B}(\textbf{r})$, see Refs. 6,7. In the relativistic
 case considered above the velocity (2) commutes with $\textbf{B}(\textbf{r})$ and the
 symmetrization is not needed.

 Clearly, the complete Dirac equation contains the effects of the
 spin-orbit interaction. For example, the exact energies obtained
 for the hydrogen atom exhibit the spin splittings which can be
 interpreted in terms of this interaction. But these are the
 energies, not the force.

 Thus the spin transverse force claimed in Ref. 1 is due to an
 approximate character of the employed Hamiltonian. If one could
 expand the Dirac Hamiltonian to high powers of $v/c$ terms,
 the result for the force would asymptotically reduce to Eq. (7).
 Such an expansion is easy to write for the presence of a magnetic
 field alone, as outlined in Ref. 2. It is much more complicated
 when both potentials are present, see [8]. However, there is no
 need to consider such expansions, the force given by Eq. (7) is
 simple, elegant and exact.

Ref. 1 also discusses possible effects related to the spin
transverse force in semiconductors. In the standard band theory of
such materials one usually employs the Schroedinger Hamiltonian
with a periodic potential plus the spin-orbit term. But again, in
principle one should use the complete Dirac Hamiltonian and then
the reasoning presented above holds. Thus one obtains the force in
the Lorentz form (7) with the electric field given as a gradient
of the periodic and external potentials. In other words, also in
this case the spin transverse force does not appear.

To summarize, we demonstrated that the spin transverse force
claimed in Ref. 1 for a relativistic electron does not exist. It
is an artefact resulting from an approximate character of the
employed Hamiltonian.

Acknowledgements: I acknowledge helpful discussions with Dr. P.
Pfeffer and Dr. T.M. Rusin.

\end{document}